\documentclass[aps,prd,twocolumn,floats,floatfix,nofootinbib]{revtex4-1}
\usepackage{graphicx}
\usepackage{amssymb}
\usepackage{epstopdf}
\usepackage{hyperref}

\usepackage{amsmath}
\usepackage{bbold}
\usepackage{color}

\begin{document}

\title{Constraining the scalar septet model through vector boson scattering}

\author{Mary-Jean Harris}

\author{Heather E.~Logan}
\email{logan@physics.carleton.ca} 

\affiliation{Ottawa-Carleton Institute for Physics, Carleton University, 1125 Colonel By Drive, Ottawa, Ontario K1S 5B6, Canada}

\date{March 10, 2017}                                 

\begin{abstract}
The scalar septet model extends the Standard Model Higgs sector by an isospin septet with hypercharge chosen to preserve $\rho \equiv M_W^2/M_Z^2 \cos^2\theta_W = 1$.  In this paper we constrain the model at high septet masses using perturbative unitarity of longitudinal vector boson scattering amplitudes.  We also apply the constraints from LHC searches for doubly-charged Higgs bosons produced in vector boson fusion, which constrain the model at lower septet masses.
We point out some important differences between the septet model and extended Higgs models that preserve custodial symmetry in the scalar spectrum.
\end{abstract}

\maketitle


\section{Introduction}

The generation of fermion and gauge boson masses in the Standard Model (SM) of particle physics requires the existence of a Higgs boson, which was discovered in 2012 at the Large Hadron Collider (LHC) at CERN~\cite{Aad:2012tfa}. However, it is possible that there are additional Higgs bosons, as described by extended Higgs models. Such additional Higgs bosons typically have different masses than that of the SM Higgs and can have nonzero electric charge. 

Arguably the most important constraint in the exploration of extended Higgs sectors is Veltman's $\rho$ parameter~\cite{Ross:1975fq}, defined in terms of the ratio of strengths of the low-energy neutral and charged weak currents, which has been measured with great precision to be~\cite{Olive:2016xmw}
\begin{equation}
	\rho = \frac{M_W^2}{M_Z^2 \cos^2\theta_W} = 1.00037 \pm 0.00023,
\end{equation}
where $M_W$ and $M_Z$ are the $W$ and $Z$ boson masses, respectively, and $\theta_W$ is the weak mixing angle.
For an extended Higgs sector containing complex scalars $\varphi_i$ with isospin $T_i$, hypercharge\footnote{We normalize the hypercharge such that $Q = T^3 + Y$, with $Q$ being the electric charge and $T^3$ being the third component of isospin.} $Y_i$, and vacuum expectation value (vev) $\langle \varphi_i^0 \rangle = v_i/\sqrt{2}$, the $\rho$ parameter at tree level is given by (see, e.g., Ref.~\cite{Gunion:1989we}),
\begin{equation}
	\rho = \frac{\sum_i [T_i (T_i + 1) - Y_i^2] v_i^2}{\sum_j 2 Y_j^2 v_j^2}.
\end{equation}
The usual SM Higgs doublet with $Y = 1/2$ yields $\rho = 1$, as favored by experiment.  The next-largest single representation that gives $\rho=1$ is the septet ($T = 3$) with $Y = 2$.  Higher-isospin solutions violate perturbative unitarity of \emph{transverse} vector boson scattering due to their excessively large weak charges~\cite{Hally:2012pu}, leaving the septet as the only perturbative exotic Higgs-sector extension in this class.

The scalar septet model, which contains such a septet in addition to the usual SM Higgs doublet, was first studied in Refs.~\cite{Hisano:2013sn,Kanemura:2013mc}.  A challenge of this model is that, at the renormalizable level, it preserves an accidental global U(1) symmetry in the scalar potential that is broken when the septet obtains a vev, leading to a physical massless Goldstone boson that couples to fermions and mediates long-range forces.  This was solved in Ref.~\cite{Hisano:2013sn} by introducing a dimension-seven effective operator coupling one septet to six Higgs doublet fields, which can be generated at the one-loop level in a fairly simple ultraviolet completion.  Another possibility is to gauge the extra U(1) in order to absorb the would-be Goldstone boson into the longitudinal component of a new massive gauge boson~\cite{Kanemura:2013mc}.  

The first general survey of the phenomenology of the scalar septet model was performed in Ref.~\cite{Alvarado:2014jva}, which studied constraints from the oblique $S$ and $T$ parameters, the couplings of the SM-like Higgs boson, and LHC searches for new physics, which together constrain the additional Higgs bosons (assumed to be roughly degenerate) to lie above about 400~GeV.  Reference~\cite{Alvarado:2014jva} also found a quite stringent constraint on the septet vev based on the LHC measurements of the couplings of the SM-like Higgs boson, which in our notation will correspond to $s_7 \lesssim 0.14$.  
The one-loop renormalization group equations for the dimension-four part of the scalar potential of the scalar septet model were computed in Ref.~\cite{Hamada:2015bra}, which showed that the model develops a Landau pole at or below $1.1 \times 10^{5} (m_7/100~{\rm GeV})^{1.10}$~GeV, where $m_7$ is the mass of the septet scalars (assuming negligible mass splittings).

Extensions of the scalar septet model have been considered in Ref.~\cite{Geng:2014oea}, which extended the analysis of the oblique parameters to include additional scalar fields besides the doublet and septet, and Ref.~\cite{Nomura:2016jnl}, which extended the model to accommodate radiative neutrino mass generation.

Our main objective in this paper is to elucidate the role of the additional Higgs bosons in unitarizing longitudinal vector boson scattering amplitudes.  This has been well-studied in the context of Higgs-sector extensions containing isospin-triplet or larger representations that preserve custodial SU(2) symmetry in the scalar sector; in such models perturbative unitarity is preserved through an interplay among the couplings of the custodial singlet(s) and a custodial fiveplet ($H_5^{++}, H_5^+, H_5^0, H_5^-, H_5^{--}$) that is always present in such models~\cite{Gunion:1989we,Falkowski:2012vh,Grinstein:2013fia,Bellazzini:2014waa}.  The best-known example is the Georgi-Machacek model~\cite{Georgi:1985nv,Chanowitz:1985ug}.  In such custodial SU(2)-preserving models, the perturbative unitarity of longitudinal vector boson scattering amplitudes can be exploited to constrain the couplings of the custodial fiveplet scalars, and hence the vevs of the higher-isospin scalars, for high masses around the TeV scale~\cite{Logan:2015xpa}.

While the scalar septet model preserves $\rho = 1$, its scalar spectrum does not exhibit custodial symmetry.  Nevertheless, we will show that longitudinal vector boson scattering is unitarized by the charged scalars in a similar way to that in the Georgi-Machacek model, and use the finite parts of the scattering amplitudes to set an upper bound on the septet vev at high septet masses.  We will also use the couplings of the septet states to constrain the septet vev using LHC searches for doubly-charged scalars in vector boson fusion~\cite{Khachatryan:2014sta}.

This paper is organized as follows.  In Sec.~\ref{sec:model} we briefly review the scalar septet model and summarize the relevant couplings of scalars to vector boson pairs.  In Sec.~\ref{sec:vv2vv} we compute the longitudinal vector boson scattering amplitudes involving scalar exchanges in the high-energy limit.  We illustrate the resulting unitarity sum rules in the scalar septet model and obtain a perturbative unitarity constraint on the septet vev at high mass.  We also point out the differences between the septet model and models with custodial symmetry in the scalar sector, using the Georgi-Machacek model as an example.  In Sec.~\ref{sec:expt} we apply the LHC limit on doubly-charged Higgs boson production in vector boson fusion~\cite{Khachatryan:2014sta} and illustrate its interplay with other constraints.  We conclude in Sec.~\ref{sec:conclusions}.  The SU(2) generators for the septet representation and some details of the coupled-channel analysis of vector boson scattering in the septet model are given in the appendices.


\section{The scalar septet model}
\label{sec:model}

\subsection{Field content}

The scalar septet model~\cite{Hisano:2013sn,Kanemura:2013mc} contains the usual complex Higgs doublet $\Phi$ together with a complex septet $X$ with isospin $T = 3$ and hypercharge $Y = 2$.  The doublet is responsible for the fermion masses as in the SM.  The fields are given by 
\begin{equation}
	{\Phi} = \begin{pmatrix} {\phi}^+ \\ {\phi}^0 \\ \end{pmatrix} , \qquad
	X = \begin{pmatrix} {\chi}^{+5} \\{\chi}^{+4} \\{\chi}^{+3} \\{\chi}^{+2} \\{\chi}^{+1} \\{\chi}^0 \\{\chi}^{-1} \\ \end{pmatrix}.
\end{equation} 
Note that the septet $X$ contains two singly-charged fields, $\chi^{+1}$ and $\chi^{-1}$, which have different $T^3$ quantum numbers and are not each others' antiparticles.  

The neutral fields can be re-expressed in terms of their vacuum expectation values (vevs) as
\begin{eqnarray}
	\phi^0 &=& (v_{\phi} + \phi^{0,r} + i \phi^{0,i})/\sqrt{2}, \nonumber \\
	\chi^0 &=& (v_{\chi} + \chi^{0,r} + i \chi^{0,i})/\sqrt{2},
\end{eqnarray} 
where ${\phi}^{0,r} $ is the real component and ${\phi}^{0,i} $ is the imaginary component of ${\phi}^0$, and the same for ${\chi}^0$.  The vevs are constrained by the $W$ and $Z$ masses to satisfy the relation
\begin{equation}
	v_{\phi}^2 + 16 v_{\chi}^2 \equiv v^2 = \frac{1}{\sqrt{2} G_F} \simeq (246~{\rm GeV})^2,
\end{equation}
where $G_F$ is the Fermi constant and $v$ is the SM Higgs vev.

\subsection{Physical spectrum}

After electroweak symmetry breaking, the model contains 15 physical scalars: two CP-even neutral scalars $h$ and $H$, a CP-odd neutral scalar $A$, two singly-charged scalar pairs $H_1^{\pm}$ and $H_2^{\pm}$, and the remaining higher-charged components of the septet $\chi^{+2}$, $\chi^{+3}$, $\chi^{+4}$, $\chi^{+5}$ and their antiparticles.  For convenience, we will also denote $\chi^{+2}$ as $H^{++}$ and its antiparticle as $H^{--}$.

The Goldstone bosons are given by\footnote{An easy way to find the Goldstone bosons is to identify the linear combinations of fields that participate in the $Z_{\mu} \partial^{\mu} \varphi$ and $W_{\mu} \partial^{\mu} \varphi$ terms in the gauge-kinetic Lagrangian.}
\begin{eqnarray}
	G^0 &=& c_7 \phi^{0,i} + s_7 \chi^{0,i}, \nonumber \\
	G^+ &=& c_7 \phi^+ + s_7 \left( \sqrt{\frac{5}{8}} \chi^{+1} - \sqrt{\frac{3}{8}} (\chi^{-1})^* \right),
\end{eqnarray}
where the mixing angle is defined by the ratio of vevs as
\begin{equation}
	c_7 \equiv \cos \theta_7 = \frac{v_{\phi}}{v}, \qquad
	s_7 \equiv \sin \theta_7 = \frac{4 v_{\chi}}{v}.
	\label{eq:c7}
\end{equation}
In particular, $s_7^2$ is the fraction of $M_W^2$ and $M_Z^2$ that is generated by the septet.

We define a mixing angle $\alpha$ between the two CP-even neutral scalars, so that
\begin{eqnarray}
	h &=& c_{\alpha} \phi^{0,r} - s_{\alpha} \chi^{0,r}, \nonumber \\
	H &=& s_{\alpha} \phi^{0,r} + c_{\alpha} \chi^{0,r},
	\label{eq:hHdef}
\end{eqnarray}
with $c_{\alpha} \equiv \cos\alpha$ and $s_{\alpha} \equiv \sin\alpha$ and $h$ defined as the lighter state.
The CP-odd neutral scalar is the state orthogonal to $G^0$,
\begin{equation}
	A = -s_7 \phi^{0,i} + c_7 \chi^{0,i}.
\end{equation}

The two singly-charged scalars can be defined as follows.  First, the state orthogonal to the charged Goldstone boson is
\begin{equation}
	H_f^+ = -s_7 \phi^+ + c_7 \left( \sqrt{\frac{5}{8}} \chi^{+1} - \sqrt{\frac{3}{8}} (\chi^{-1})^* \right),
\end{equation}
where the $f$ subscript refers to the fact that this linear combination of fields couples to fermion pairs but not to gauge boson pairs.  The remaining orthogonal combination consists entirely of septet states,
\begin{equation}
	H_V^+ = \sqrt{\frac{3}{8}} \chi^{+1} + \sqrt{\frac{5}{8}} (\chi^{-1})^*,
\end{equation}
where the $V$ subscript refers to the fact that this linear combination couples to gauge boson pairs but not to fermion pairs.
We write the physical singly-charged mass eigenstates as linear combinations of these according to
\begin{eqnarray}
	H_1^+ &=& c_{\gamma} H_f^+ - s_{\gamma} H_V^+, \nonumber \\
	H_2^+ &=& s_{\gamma} H_f^+ + c_{\gamma} H_V^+,
\end{eqnarray}
with $H_1^+$ being the lighter state and $c_{\gamma} \equiv \cos\gamma$, $s_{\gamma} \equiv \sin\gamma$.

Finally we define
\begin{equation}
	H^{++} = \chi^{+2}.
\end{equation}
This state couples to gauge boson pairs, but not to fermions.

\subsection{Couplings to gauge boson pairs}

The couplings of the scalars to pairs of gauge bosons arise from the gauge-kinetic terms in the Lagrangian,
\begin{equation}
	\mathcal{L} \supset (\mathcal{D}_{\mu}{\Phi})^\dagger(\mathcal{D}^{\mu}{\Phi})
	+ (\mathcal{D}_{\mu} X)^{\dagger} (\mathcal{D}^{\mu} X),
	\label{gaugekinetic}
\end{equation}
where the covariant derivative is given by
\begin{eqnarray}
	\mathcal{D}_{\mu} &=& \partial_{\mu} - i\frac{g}{\sqrt{2}}(W^+_{\mu}T^++ W^-_{\mu}T^-) 
	\nonumber \\
	&& -\frac{ie}{s_Wc_W}Z_{\mu}(T^3 -s_W^2Q) 
	 -ieA_{\mu}Q. 
\end{eqnarray}
Here $T^{\pm} = T^1 \pm i T^2$.  The generators for the doublet representation are $T^a = \sigma^a/2$ with $\sigma^a$ being the Pauli matrices.  The generators for the septet representation are given in Appendix~\ref{app:generators}.

Only $h$, $H$, $H_V^{\pm}$, and $H^{\pm\pm}$ participate in three-point couplings to pairs of gauge bosons.  In each case the Feynman rule for the three-point vertex can be written as
\begin{equation}
	\varphi V_1^{\mu} V_2^{\nu}: \ \ \ \frac{2 i M_{V_1} M_{V_2}}{v} g^{\mu\nu} \kappa^{\varphi}_{V_1V_2},
\end{equation}
where for each coupling we define a relative coupling factor $\kappa^{\varphi}_{V_1V_2}$.  In this notation, the SM Higgs coupling factors are $\kappa^{h_{\rm SM}}_{WW} = \kappa^{h_{\rm SM}}_{ZZ} = 1$.
The relative coupling factors of the scalars in the septet model are given by
\begin{eqnarray}
	\kappa^h_{WW} = \kappa^h_{ZZ} &=& c_7 c_{\alpha} - 4 s_7 s_{\alpha} \equiv \kappa_V^h, \nonumber \\
	\kappa^H_{WW} = \kappa^H_{ZZ} &=& c_7 s_{\alpha} + 4 s_7 c_{\alpha} \equiv \kappa_V^H, \nonumber \\
	\kappa^{H_V^+}_{WZ} &=& - \sqrt{15} s_7, \nonumber \\
	\kappa^{H^{++}}_{WW} &=& \sqrt{15} s_7.
	\label{eq:coups}
\end{eqnarray}
The couplings of the singly-charged Higgs mass eigenstates are modified by the mixing angle $\gamma$, yielding
\begin{equation}
	\kappa^{H_1^+}_{WZ} = \sqrt{15} s_7 s_{\gamma}, \qquad
	\kappa^{H_2^+}_{WZ} = -\sqrt{15} s_7 c_{\gamma}.
\end{equation}

Note that setting the septet vev to zero corresponds to $s_7 = 0$, in which case the couplings of the singly- and doubly-charged scalars to gauge boson pairs vanish.  Further taking the doublet-septet mixing angle $\alpha$ to zero, the couplings of $h$ to gauge boson pairs become identical to those of the SM Higgs and the couplings of $H$ to gauge boson pairs vanish.

\section{Longitudinal vector boson scattering}
\label{sec:vv2vv}

\subsection{Standard Model}

To set the stage, we first review the Higgs-exchange contributions to longitudinal vector boson scattering in the SM in the high energy limit.  In what follows we will express the matrix elements in terms of the Mandelstam variables $s$, $t$, and $u$, and neglect terms of order $M_V^2$ ($V = W, Z$) relative to the center-of-mass energy of the scattering process.

The matrix element for longitudinal $W^+W^- \to W^+W^-$ scattering receives contributions from $s$-channel and $t$-channel Higgs exchange which are given by
\begin{eqnarray}
	\mathcal{M} 
	&=& -\frac{1}{v^2} \left(\frac{s^2}{s-m_h^2} + \frac{t^2}{t-m_h^2} \right) \nonumber \\
	&\simeq& -\frac{1}{v^2}(s + t + 2m_h^2),
	\label{eq:MSMww}
\end{eqnarray}
where in the second line we have expanded the propagators in powers of $m_h^2/s$ and $m_h^2/t$ and dropped terms suppressed by powers of the scattering energy.  
The first two terms in the parentheses grow with the square of the scattering energy and are canceled by the diagrams involving only gauge bosons.  The last term gives rise to the famous upper bound on the SM Higgs boson mass of Lee, Quigg and Thacker~\cite{Lee:1977yc}, as follows.  

The matrix element for $2\to 2$ scattering can be expanded in terms of the partial wave amplitudes $a_J$ according to
\begin{equation}
	\mathcal{M} = 16 \pi \sum_{J =0}^{\infty} (2J + 1) a_J P_J(\cos\theta),
\end{equation}
where $J$ is the orbital angular momentum of the final state and $P_J(\cos\theta)$ is the corresponding Legendre polynomial.  Tree-level partial wave unitarity dictates that
\begin{equation}
	|{\rm Re} \, a_0| \leq 1/2.
\end{equation}
Applying this to the last term in Eq.~(\ref{eq:MSMww}) leads to an upper bound on the SM Higgs boson mass,
\begin{equation}
	m_h^2 \leq 4 \pi v^2 \simeq (872~{\rm GeV})^2.
\end{equation}

The matrix element for longitudinal $W^{\pm}Z \to W^{\pm}Z$ scattering receives a contribution from $t$-channel Higgs exchange, which yields
\begin{eqnarray}
	\mathcal{M} &=& -\frac{1}{v^2} \left( \frac{t^2}{t - m_h^2} \right) \nonumber \\
	&\simeq& -\frac{1}{v^2} (t + m_h^2),
	\label{eq:SMWZ}
\end{eqnarray}
where again we have expanded the propagator in the second line.  The term proportional to $t$ is canceled by the diagrams involving only gauge bosons, while the corresponding limit on the Higgs mass from perturbative unitarity is
\begin{equation}
	m_h^2 \leq 8 \pi v^2 \simeq (1230~{\rm GeV})^2.
\end{equation}

The matrix element for longitudinal $ZZ \to ZZ$ scattering receives contributions from $s$-, $t$- and $u$-channel Higgs exchange, which give
\begin{eqnarray}
	\mathcal{M} &=& - \frac{1}{2v^2} \left( \frac{s^2}{s - m_h^2} + \frac{t^2}{t - m_h^2} + \frac{u^2}{u - m_h^2} \right) \nonumber \\
	&\simeq& -\frac{3 m_h^2}{2v^2},
\end{eqnarray}
where we have again expanded the propagators and used $s + t + u \simeq 0$ in the second line, thereby consistently neglecting $M_Z^2$ in the high energy limit.  In this expression we have incorporated a factor of $1/\sqrt{2}$ for each pair of identical particles in the initial and final states.  Note that the Higgs-exchange diagrams for $ZZ \to ZZ$ scattering by themselves do not grow with scattering energy; this is required because there are no $ZZ \to ZZ$ scattering diagrams involving only gauge bosons.  The corresponding limit on the Higgs mass from perturbative unitarity is
\begin{equation}
	m_h^2 \leq \frac{16 \pi v^2}{3} \simeq (1010~{\rm GeV})^2.
\end{equation}

These processes were combined in a coupled channel analysis in Ref.~\cite{Lee:1977yc} to obtain a slightly more stringent upper bound on the SM Higgs mass.

\subsection{Scalar septet model}
\label{sec:septet-unitarity}

We now consider each of these longitudinal gauge boson scattering processes in the scalar septet model.  By matching the parts of the amplitudes that grow with the scattering energy to the corresponding parts of the SM Higgs amplitudes, we illustrate two coupling sum rules that hold in the scalar septet model.  We then use the piece of the amplitude that does not grow with scattering energy to derive new constraints on the scalar couplings and masses from perturbative unitarity.

The matrix element for longitudinal $W^+W^- \to W^+ W^-$ scattering in the scalar septet model receives contributions from $s$- and $t$-channel exchange of the neutral scalars $h$ and $H$ and $u$-channel exchange of the doubly-charged scalar $H^{++}$, which yield
\begin{eqnarray}
	\mathcal{M} &=& -\frac{1}{v^2} \left[ (\kappa_V^h)^2 \left( \frac{s^2}{s - m_h^2} + \frac{t^2}{t - m_h^2} \right) \right. \nonumber \\
	&& \left. + (\kappa_V^H)^2 \left( \frac{s^2}{s - m_H^2} + \frac{t^2}{t - m_H^2} \right) \right. \nonumber \\ 
	&& \left. + (\kappa_{WW}^{H^{++}})^2 \left( \frac{u^2}{u - m_{H^{++}}^2} \right) \right] \nonumber \\
	&\simeq& - \frac{1}{v^2} \left[ \left( (\kappa_V^h)^2 + (\kappa_V^H)^2 - (\kappa_{WW}^{H^{++}})^2 \right) (s + t) \right. \nonumber \\
	&& \left. + 2 (\kappa_V^h)^2 m_h^2 + 2 (\kappa_V^H)^2 m_H^2 + (\kappa_{WW}^{H^{++}})^2 m_{H^{++}}^2 \right],
\end{eqnarray}
where we have used $u \simeq -(s + t)$ in the high energy limit.

Comparing this expression to Eq.~(\ref{eq:MSMww}), we obtain the sum rule,
\begin{equation}
	(\kappa_V^h)^2 + (\kappa_V^H)^2 - (\kappa_{WW}^{H^{++}})^2 = 1,
	\label{eq:WWsumrule}
\end{equation}
which of course is satisfied by the couplings of Eq.~(\ref{eq:coups}).  We also obtain a constraint on the scalar masses and couplings,
\begin{equation}
	2 ({\kappa}_V^h)^2 m_h^2  + 2 ({\kappa}_V^H)^2 m_H^2 + ({\kappa}_{WW}^{H^{++}})^2 m_{H^{++}}^2 \leq 8 \pi v^2.
	\label{eq:WWmax}
\end{equation}

Equations~(\ref{eq:WWsumrule}) and (\ref{eq:WWmax}) can be exploited as follows.  First, for fixed $s_7$, the maximum value of $\kappa^h_V$ is obtained when $\kappa^H_V = 0$:
\begin{equation}
	\left. ({\kappa}_V^h)^2 \right|_{\rm max} = 1 + ({\kappa}_{WW}^{H^{++}})^2.
	\label{eq:khmax}
\end{equation}
Inserting this into Eq.~(\ref{eq:WWmax}) with $\kappa_{WW}^H = 0$ yields the most conservative upper bound on the $H^{++}$ coupling to $W^+W^+$ as a function of its mass,
\begin{equation}
	({\kappa}_{WW}^{H^{++}})^2 \leq \frac{8{\pi}v^2 - 2m_h^2}{m^2_{H^{++}} + 2m_h^2}.
	\label{eq:kwwmax}
\end{equation}
This in turn sets an upper bound on $s_7$ that falls with increasing $m_{H^{++}}$,
\begin{equation}
	s_7^2 \leq \frac{1}{15} \frac{8{\pi}v^2 - 2m_h^2}{m^2_{H^{++}} + 2 m_h^2}
	\simeq \left( \frac{315~{\rm GeV}}{m_{H^{++}}} \right)^2,
	\label{eq:s7boundWW}
\end{equation}
where the last expression holds for $m_{H^{++}}^2 \gg 2 m_h^2$.  

The matrix element for longitudinal $W^{\pm}Z \to W^{\pm}Z$ scattering receives contributions from $t$-channel exchange of $h$ and $H$ as well as $s$- and $u$-channel exchange of $H_1^+$ and $H_2^+$, which yield
\begin{eqnarray}
	\mathcal{M} &=& - \frac{1}{v^2} \left[ (\kappa_V^h)^2 \left( \frac{t^2}{t - m_h^2} \right) 
	+ (\kappa_V^H)^2 \left( \frac{t^2}{t - m_H^2} \right) \right. \nonumber \\
	&& + \left. (\kappa_{WZ}^{H_1^+})^2 \left( \frac{s^2}{s - m_{H_1^+}^2} + \frac{u^2}{u - m_{H_1^+}^2} \right) \right. \nonumber \\
	&& + \left. (\kappa_{WZ}^{H_2^+})^2 \left( \frac{s^2}{s - m_{H_2^+}^2} + \frac{u^2}{u - m_{H_2^+}^2} \right) \right] \nonumber \\
	&\simeq& -\frac{1}{v^2} \left[ \left( (\kappa_V^h)^2 + (\kappa_V^H)^2 - (\kappa_{WZ}^{H_1^+})^2 - (\kappa_{WZ}^{H_2^+})^2 \right) t \right. \nonumber \\
	&& + \left. (\kappa_V^h)^2 m_h^2 + (\kappa_V^H)^2 m_H^2 + 2 (\kappa_{WZ}^{H_1^+})^2 m_{H_1^+}^2 \right. \nonumber \\
	&& \left. + 2 (\kappa_{WZ}^{H_2^+})^2 m_{H_2^+}^2 \right],
\end{eqnarray}
where we have used $s + u \simeq -t$.

Comparing this expression to Eq.~(\ref{eq:SMWZ}), we obtain the sum rule,
\begin{equation}
	(\kappa_V^h)^2 + (\kappa_V^H)^2 - (\kappa_{WZ}^{H_1^+})^2 - (\kappa_{WZ}^{H_2^+})^2 = 1,
	\label{eq:H+sumrule}
\end{equation}
which of course is satisfied by the couplings of Eq.~(\ref{eq:coups}).  We also obtain a constraint on the masses and couplings,
\begin{eqnarray}
	(\kappa_V^h)^2 m_h^2 + (\kappa_V^H)^2 m_H^2 + 2 (\kappa_{WZ}^{H_1^+})^2 m_{H_1^+}^2 && \nonumber \\
	+ 2 (\kappa_{WZ}^{H_2^+})^2 m_{H_2^+}^2 &\leq& 8 \pi v^2.
	\label{eq:H+bound}
\end{eqnarray}

We can exploit Eqs.~(\ref{eq:H+sumrule}) and (\ref{eq:H+bound}) to set an upper bound on $s_7$ as a function of the $H_1^+$ and $H_2^+$ masses.  Again taking $\kappa_V^H = 0$ to obtain the most conservative bound, we find
\begin{eqnarray}
	s_7^2 &\leq& \frac{1}{15} \frac{8 \pi v^2 - m_h^2}{2 s_{\gamma}^2 m_{H_1^+}^2 + 2 c_{\gamma}^2 m_{H_2^+}^2 + m_h^2} \nonumber \\
	&\simeq& \frac{(225~{\rm GeV})^2}{(s_{\gamma}^2 m_{H_1^+}^2 + c_{\gamma}^2 m_{H_2^+}^2)}.
	\label{eq:WZunitarity}
\end{eqnarray}
For common charged-Higgs masses $m_{H^{++}} \simeq m_{H_1^+} \simeq m_{H_2^+}$, this constraint is more stringent than the one from $W^+W^- \to W^+W^-$ scattering found in Eq.~(\ref{eq:s7boundWW}).  A coupled-channel analysis taking all septet states to be degenerate in mass yields an identical constraint in the high charged-Higgs mass limit; see Appendix~\ref{app:coupled} for details.

Finally, the matrix element for longitudinal $ZZ \to ZZ$ scattering receives contributions from $s$-, $t$- and $u$-channel exchange of $h$ and $H$, which yield
\begin{eqnarray}
	\mathcal{M} &=& - \frac{1}{2 v^2} \left[ (\kappa^h_V)^2 \left( \frac{s^2}{s - m_h^2} + \frac{t^2}{t - m_h^2} + \frac{u^2}{u - m_h^2} \right) \right. \nonumber \\
	&& \left. + (\kappa^H_V)^2 \left( \frac{s^2}{s - m_H^2} + \frac{t^2}{t - m_H^2} + \frac{u^2}{u - m_H^2} \right) \right] \nonumber \\
	&\simeq& - \frac{3}{2 v^2} \left[ (\kappa^h_V)^2 m_h^2 + (\kappa^H_V)^2 m_H^2 \right],
\end{eqnarray}
where we have again incorporated a factor of $1/\sqrt{2}$ for each pair of identical particles in the initial and final states.  There is no sum rule here because the terms that grow with the collision energy have already dropped out.  We do obtain a constraint on the masses and couplings,
\begin{equation}
	(\kappa^h_V)^2 m_h^2 + (\kappa^H_V)^2 m_H^2 \leq \frac{16 \pi v^2}{3}.
\end{equation}

\subsection{Comparison to models with custodial symmetry}
\label{sec:custodial}

In order to highlight the unique features of the scalar septet model, we now compare the sum rules and mass and coupling constraints obtained from longitudinal vector boson scattering in the septet model to those in models with custodial symmetry.  For concreteness we focus on the Georgi-Machacek model~\cite{Georgi:1985nv,Chanowitz:1985ug}, but entirely analogous expressions appear in generalizations of the Georgi-Machacek model, as described in detail in Sec.~IV~B of Ref.~\cite{Logan:2015xpa}.

The scalars in the Georgi-Machacek model that couple to vector boson pairs are the custodial singlets $h$ and $H$ and the states of the custodial fiveplet ($H_5^{++}, H_5^+, H_5^0, H_5^-, H_5^{--}$), which are constrained by the custodial symmetry to have a common mass $m_5$.  We note already three differences compared to the septet model: the presence of a third neutral scalar $H_5^0$ that couples to vector boson pairs; the mass-degeneracy of the singly- and doubly-charged scalars; and the absence of mixing between the custodial-fiveplet singly-charged scalar that couples to vector boson pairs and the additional singly-charged scalar (part of a custodial triplet in the Georgi-Machacek model) that couples to fermion pairs.

The relative coupling factors of the scalars in the Georgi-Machacek model that couple to vector boson pairs are given by
\begin{eqnarray}
	{\kappa}_{WW}^h = {\kappa}_{ZZ}^h &=& c_{\alpha} c_H - \sqrt{\frac{8}{3}} s_{\alpha} s_H \equiv \kappa_V^h, \nonumber \\
	{\kappa}_{WW}^H = {\kappa}_{ZZ}^H &=& s_{\alpha} c_H + \sqrt{\frac{8}{3}} c_{\alpha} s_H \equiv \kappa_V^H, \nonumber \\
	{\kappa}_{WW}^{H_5^0} = \frac{1}{\sqrt{3}} s_H,  &&\qquad
	{\kappa}_{ZZ}^{H_5^0} = - \frac{2}{\sqrt{3}} s_H, \nonumber \\
	{\kappa}_{WZ}^{H_5^+} &=& - s_H, \nonumber \\
	{\kappa}_{WW}^{H_5^{++}} &=& \sqrt{2} s_H.
	\label{eq:GMcoups}
\end{eqnarray}
Here the cosines of the mixing angles $c_H$ and $c_{\alpha}$ are defined analogously to those in Eqs.~(\ref{eq:c7}) and (\ref{eq:hHdef}), so that $s_H^2 \equiv 1 - c_H^2$ is the fraction of $M_W^2$ and $M_Z^2$ that is generated by the higher-isospin multiplets and the SU(2)$_L$-doublet component of $h$ ($H$) is given by $c_{\alpha}$ ($s_{\alpha}$).

The calculation of the longitudinal vector boson scattering amplitudes goes through exactly as in the septet model, except that the neutral custodial-fiveplet state $H_5^0$ contributes along with $h$ and $H$ in all the electrically-neutral channels.  The sum rule from $W^+W^- \to W^+W^-$ scattering then becomes
\begin{equation}
	({\kappa}_V^h)^2 + ({\kappa}_V^H)^2 + ({\kappa}_{WW}^{H_5^0})^2 - (\kappa_{WW}^{H_5^{++}})^2 = 1.
\end{equation}
In fact, because the custodial symmetry enforces the relationship $\kappa_{WW}^{H_5^0} = \kappa_{WW}^{H_5^{++}}/\sqrt{6}$ [see Eq.~(\ref{eq:GMcoups})], we can simplify this to read
\begin{equation}
	({\kappa}_V^h)^2 + ({\kappa}_V^H)^2 - \frac{5}{6} ({\kappa}_{WW}^{H_5^{++}})^2 = 1.
\end{equation}
Note the extra factor of $5/6$ on the last term compared to the corresponding sum rule for the septet model in Eq.~(\ref{eq:WWsumrule}).  

The constraint on the masses and couplings from $W^+W^- \to W^+W^-$ scattering is also affected by the presence of $H_5^0$, yielding
\begin{eqnarray}
	2 ({\kappa}_V^h)^2 m_h^2 + 2 ({\kappa}_V^H)^2 m_H^2 + 2 ({\kappa}_{WW}^{H_5^0})^2 m_5^2  && \nonumber \\
	 + ({\kappa}_{WW}^{H_5^{++}})^2 m_5^2 &\leq& 8{\pi}v^2,
\end{eqnarray}
which can be condensed using the relationship among custodial-fiveplet couplings to read
\begin{equation}
	2 ({\kappa}_V^h)^2 m_h^2 + 2 ({\kappa}_V^H)^2 m_H^2 + \frac{4}{3} ({\kappa}_{WW}^{H_5^{++}})^2 m_5^2 \leq 8{\pi}v^2.
\end{equation}
Note the extra factor of $4/3$ on the last term compared to the corresponding constraint on the doubly-charged scalar of the septet model in Eq.~(\ref{eq:WWmax}).

These extra factors lead to modified versions of Eqs.~(\ref{eq:khmax}) and (\ref{eq:kwwmax}):
\begin{equation}
	\left. ({\kappa}_V^h)^2 \right|_{\rm max} = 1 + \frac{5}{6}({\kappa}_{WW}^{H_5^{++}})^2,
\end{equation}
and
\begin{equation}
	({\kappa}_{WW}^{H_5^{++}})^2 \leq \frac{3(8{\pi}v^2 - 2m_h^2)}{4m^2_5 + 5 m_h^2},
\end{equation}
implying slightly tighter upper bounds on $\kappa_V^h$ for fixed $\kappa_{WW}^{H_5^{++}}$ and on $\kappa_{WW}^{H_5^{++}}$ for fixed $m_5 \gg m_h$ than in the septet model.

The fact that the custodial-fiveplet states in the Georgi-Machacek model all have the same mass allows the constraints from different vector boson scattering processes to be combined in a coupled channel analysis, leading to still tighter constraints.  A detailed analysis has been performed in Ref.~\cite{Logan:2015xpa}, which found an upper bound from perturbative unitarity of
\begin{equation}
	({\kappa}_{WW}^{H_5^{++}})^2 \leq \frac{6}{5} \frac{(16{\pi}v^2 - 5m_h^2)}{(4m_5^2 + 5 m_h^2)}.
\end{equation}
For fixed doubly-charged scalar mass and neglecting $m_h$, this coupled-channel bound on ${\kappa}_{WW}^{H_5^{++}}$ is $\sqrt{6/5}$ times the corresponding bound on $\kappa_{WW}^{H^{++}}$ in the septet model -- i.e., the perturbative unitarity bound on this coupling is about 10\% stronger in the septet model than in the Georgi-Machacek model.

For completeness, we give the sum rules and constraints on masses and couplings in the Georgi-Machacek model from the remaining individual scattering channels.  From $W^{\pm} Z \to W^{\pm} Z$ scattering, the sum rule reads
\begin{equation}
	({\kappa}_V^h)^2 + ({\kappa}_V^H)^2 + {\kappa}_{WW}^{H_5^0} {\kappa}_{ZZ}^{H_5^0} - ({\kappa}_{WZ}^{H_5^+})^2 = 1,
\end{equation}
and the perturbative unitarity constraint on masses and couplings is
\begin{eqnarray}
	({\kappa}_V^h)^2 m_h^2 + ({\kappa}_V^H)^2 m_H^2 + {\kappa}_{WW}^{H_5^0} {\kappa}_{ZZ}^{H_5^0} m_5^2 && \nonumber \\
	+ 2 ({\kappa}_{WZ}^{H_5^+})^2 m_5^2 &\leq& 8{\pi}v^2.
\end{eqnarray}
From $ZZ \to ZZ$ scattering, the perturbative unitarity constraint on masses and couplings is
\begin{equation}
	({\kappa}_V^h)^2 m_h^2 + ({\kappa}_V^H)^2 m_H^2 + ({\kappa}_{ZZ}^{H_5^0})^2 m_5^2 \leq \frac{16{\pi}v^2}{3}.
\end{equation}

\section{Constraint from charged scalar searches in vector boson fusion}
\label{sec:expt}

The LHC experiments have performed direct searches for on-shell production of doubly- and singly-charged Higgs bosons in vector boson fusion, followed by decays back to vector boson pairs.  These searches probe directly the scalar couplings constrained by the longitudinal vector boson scattering amplitudes considered in the previous section.

The following searches using LHC data have been performed to date:
\begin{enumerate}
\item[\it i.] An ATLAS measurement of the like-sign $WW$ cross section in vector boson fusion~\cite{Aad:2014zda} using the like-sign dilepton final state in 20.3~fb$^{-1}$ of proton-proton data at a centre-of-mass energy of 8~TeV.  This was recast in the context of the Georgi-Machacek model by Ref.~\cite{Chiang:2014bia} to set a limit on the size of the triplet vev (equivalently the $H_5^{++} W^- W^-$ coupling) as a function of the doubly-charged Higgs mass.
\item[\it ii.] A CMS search for doubly-charged Higgs boson production in like-sign $W$ boson fusion with decays back to $WW$~\cite{Khachatryan:2014sta} using the like-sign dilepton final state in 19.4~fb$^{-1}$ of data at 8~TeV.
\item[\it iii.] An ATLAS search for singly-charged Higgs boson production in $WZ$ fusion with decays back to $WZ$~\cite{Aad:2015nfa} with $WZ \to qq\ell \ell$ in 20.3~fb$^{-1}$ of data at 8~TeV.
\item[\it iv.] A CMS search for singly-charged Higgs boson production in $WZ$ fusion with decays back to $WZ$~\cite{CMS:2016szz} using the 3-lepton final state in 15.2~fb$^{-1}$ of data at 13~TeV.
\end{enumerate}

The most sensitive of these to constrain the septet model is the CMS search for doubly-charged Higgs boson production in Ref.~\cite{Khachatryan:2014sta}.  We translate the experimental upper bounds on the doubly-charged Higgs production cross section into bounds on $(\kappa_{WW}^{H^{++}})^2$ using the vector boson fusion cross sections computed at next-to-next-to-leading order in QCD for the 8~TeV LHC in Ref.~\cite{Zaro:2015ika}.  These were computed for the Georgi-Machacek model setting $s_H = 1$, or equivalently $(\kappa_{WW}^{H^{++}})^2 = 2$.  We then use Eq.~(\ref{eq:coups}) to translate the upper bound on $(\kappa_{WW}^{H^{++}})^2$ into an upper bound on $s_7$.

\begin{figure}
\resizebox{0.48\textwidth}{!}{\includegraphics{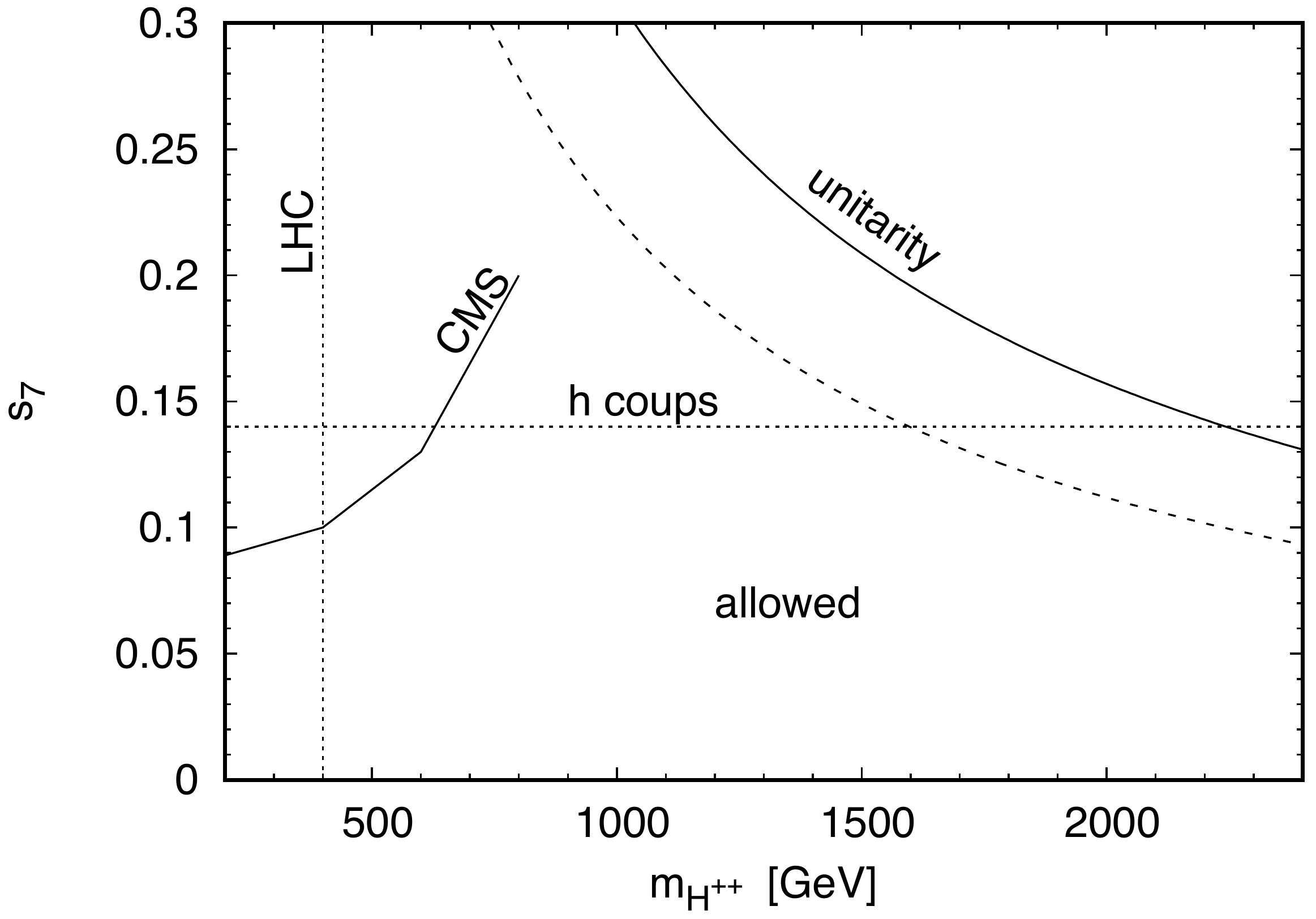}}
\caption{Constraints on $s_7$ as a function of $m_{H^{++}}$ in the scalar septet model. See text for details.}
\label{fig:s7plot}
\end{figure}

The resulting constraint is shown in Fig.~\ref{fig:s7plot} as the solid line labeled `CMS'; the region above the line is excluded.  Also shown are the lower bound of about 400~GeV on the common mass of the septet states from a combination of LHC searches (vertical short-dashed line labeled `LHC') and the upper bound of about 0.14 on $s_7$ from measurements of the couplings of $h$ (horizontal short-dashed line labeled `h coups'), both from Ref.~\cite{Alvarado:2014jva}; the region above and to the left of these lines is excluded.  Finally we show the upper bound on $s_7$ as a function of $m_{H^{++}}$ from perturbative unitarity of $WW$ scattering from Eq.~(\ref{eq:s7boundWW}) (solid line labeled `unitarity'); the region above this line is excluded.  We also show as the long-dashed line the perturbative unitarity constraint from the coupled-channel analysis that holds when $m_{H^{++}} = m_{H_1^+} = m_{H_2^+}$ [Eq.~(\ref{eq:s7coupled})].

\section{Conclusions}
\label{sec:conclusions}

In this paper we studied the constraints on the scalar septet model from perturbative unitarity of longitudinal vector boson scattering amplitudes and from LHC searches for singly- and doubly-charged Higgs bosons in vector boson fusion with decays back to vector boson pairs.  We showed that perturbative unitarity provides a significant new constraint on the septet vev for septet scalar masses above about 1600~GeV, and that the LHC searches in vector boson fusion exclude additional parameter space for doubly-charged scalar masses below 600~GeV.  Overall, the upper bound on $s_7$ due to Higgs coupling measurements from Ref.~\cite{Alvarado:2014jva} constrains the fraction of $M_W^2$ and $M_Z^2$ that can be generated by the septet vev to be below 2\%, already a very stringent constraint.  Future improvements of the LHC measurements of the SM-like Higgs boson's couplings and searches for singly- and doubly-charged Higgs boson production in vector boson fusion may further tighten these constraints, but the upper bound on the septet vev at sufficiently large septet masses will always be dominated by perturbative unitarity requirements.

\appendix

\section{SU(2) generators for the septet representation}
\label{app:generators}

The SU(2)$_L$ generators for the septet representation are given as follows, with $T^{\pm} = T^1 \pm i T^2$:
\begin{eqnarray}
T^+ &=& \left( \begin{array}{ccccccc} 0 & \sqrt{6} &0&0&0&0 & 0 \\
                                0 & 0 &\sqrt{10}&0&0&0 & 0 \\
			            0 & 0 &0&2\sqrt{3}&0&0 & 0 \\
        			      0 & 0 &0&0&2\sqrt{3}&0 & 0  \\ 
                               0 & 0 &0&0&0&\sqrt{10} & 0  \\
                               0 & 0 &0&0&0&0&\sqrt{6} \\
                               0 & 0 &0&0&0&0 &0 \end{array} \right), \nonumber \\
T^3 &=& \left( \begin{array}{ccccccc} 3 & 0 &0&0&0&0 & 0 \\
                                0 & 2 &0&0&0&0 & 0 \\
			            0 & 0 &1&0&0&0 & 0 \\
        			      0 & 0 &0&0&0&0 & 0  \\ 
                               0 & 0 &0&0&-1&0 & 0  \\
                               0 & 0 &0&0&0&-2 & 0  \\
                               0 & 0 &0&0&0&0 &-3 \end{array} \right),
\end{eqnarray}
and $T^- = (T^+)^T$.

\section{Coupled-channel analysis for the septet model}
\label{app:coupled}

If the masses of the singly- and doubly-charged scalars in the septet model are all equal, $m_{H^{++}} = m_{H_1^+} = m_{H_2^+} \equiv m_7$, then the amplitudes for the scattering processes $W^+W^- \to W^+W^-$, $W^+W^- \leftrightarrow ZZ$, and $ZZ \to ZZ$ can be combined in a coupled-channel analysis, which generally results in a more stringent constraint from perturbative unitarity.  The analysis involves finding the largest eigenvalue of the coupled-channel scattering matrix among the $W^+W^-$ and $ZZ$ final states.  

Using the matrix elements found in Sec.~\ref{sec:septet-unitarity} and incorporating the additional factor of $1/\sqrt{2}$ for the $W^+W^- \leftrightarrow ZZ$ amplitude compared to that for $W^{\pm}Z \to W^{\pm}Z$, we obtain
\begin{eqnarray}
	\lambda_+ &=& \frac{1}{2} \left\{ 15 s_7^2 (m_7^2 + 5 m_h^2) + 5 m_h^2 \right. \nonumber \\
	&& \left. + \sqrt{3} \left[ (15 s_7^2)^2 (3 m_7^4 + 2 m_7^2 m_h^2 + m_h^4) \right. \right. \nonumber \\
	&& \left. \left. + 30 s_7^2 m_h^2 (m_7^2 + m_h^2) + m_h^4 \right]^{1/2} \right\},
\end{eqnarray}
where $\lambda_+$ is the larger eigenvalue of the scattering matrix normalized in such a way that the condition for perturbative unitarity reads
\begin{equation}
	\lambda_+ \leq 8 \pi v^2.
\end{equation}
For $m_7 \gg m_h$, we can drop the $m_h$ dependence and solve for the upper bound on $s_7$,
\begin{equation}
	s_7^2 \leq \frac{4 \pi v^2}{15 m_7^2} \simeq \left( \frac{225~\rm{GeV}}{m_7} \right)^2.
	\label{eq:s7coupled}
\end{equation}
This turns out to be identical to the perturbative unitarity bound obtained from the $W^{\pm}Z \to W^{\pm}Z$ channel in Eq.~(\ref{eq:WZunitarity}) when $m_{H_1^+} = m_{H_2^+} = m_7$.

\begin{acknowledgments}
This work was partially supported by the Natural Sciences and Engineering Research Council of Canada.  H.E.L.\ is also partially supported through the grant H2020-MSCA-RISE-2014 no.\ 645722 (NonMinimalHiggs).
\end{acknowledgments}


\end{document}